\documentclass[runningheads,11pt]{llncs}

\usepackage{fullpage}
\usepackage{graphicx}

\usepackage{cite}
\usepackage{algorithm}
\usepackage{algpseudocode}
\usepackage{xcolor}
\usepackage{listings}
\usepackage{subcaption}

\usepackage{url}
\usepackage{xspace}
\usepackage{hyperref}
\hypersetup{
 pdfborder={0 0 0},
 colorlinks=true,
 linkcolor=blue,
 urlcolor=blue,
 citecolor=blue
}

\algdef{SE}[DOWHILE]{Do}{doWhile}{\algorithmicdo}[1]{\algorithmicwhile\ #1}%
\algtext*{EndFor}
\algtext*{EndIf}

\definecolor{mymauve}{rgb}{0.58,0,0.82}

\lstdefinelanguage{lustre}{
  escapechar=|,
  keywords=[1]{
    node,function,returns,let,tel,var,const,
    assert,when,current,true,false,pre,if,
    then,else,not,and,or,contract
  },
  keywords=[2]{int,real,bool},
  keywords=[3]{
    ensure, guarantee
  },
  keywords=[4]{
    require, assume
  },
  keywords=[5]{
    mode,@mode,contract,@contract,import
  },
  keywords=[6]{
    @var,@const
  },
  sensitive=true,
  morecomment=[l]{--},
  morestring=[b]",
  morestring=[b]""",
  numberstyle=\tiny\color{white!40!black},
  keywordstyle=[1]\color{blue},
  keywordstyle=[2]\color{orange}\bfseries,
  keywordstyle=[3]\color{red!60!black}\bfseries,
  keywordstyle=[4]\color{green!60!black}\bfseries,
  keywordstyle=[5]\color{orange!60!black}\bfseries,
  keywordstyle=[6]\color{blue!60!black}\bfseries,
  commentstyle=\color{white!60!black}\it\bfseries,
  stringstyle=\color{gray},
  aboveskip=1mm,
  belowskip=1mm,
  showstringspaces=false,
  numbers=left,
  numbersep=5pt,
  breaklines=true,
  breakatwhitespace=true,
  columns=fullflexible,
  keepspaces=true,
  tabsize=3,
}

\newcommand{\tinyLustreLst}[0]{
  \lstset{language=lustre, basicstyle={\footnotesize\ttfamily}}
}
\lstnewenvironment{lustreTiny}[1][0pt]{
  \tinyLustreLst
  \lstset{xleftmargin=#1} 
}{}
\lstnewenvironment{lustreTinyMath}[1][0pt]{
  \tinyLustreLst
  \lstset{mathescape, xleftmargin=#1} 
}{}

\newcommand{\code}[1]{{\small\texttt{#1}}}

\newcommand{\kind}{\textsc{Kind}~2\xspace}
\newcommand{\jkind}{\textsc{JKind}\xspace}

\newcommand{\define}[1]{{\emph{#1}}}

\newcommand{\sys}{S}
\newcommand{\s}[1]{\mathbf{s}_{#1}}
\newcommand{\y}{\mathbf{y}}
\newcommand{\z}{\mathbf{z}}
\newcommand{\nxtS}[1]{\s{#1}'}
\newcommand{\init}{I}
\newcommand{\trans}{T}
\newcommand{\sysTuple}{
  \langle \s{}, \init[\s{}], \trans[\s{}, \nxtS{}] \rangle
}
\newcommand{\sysTupleAbr}{
  \langle \s{}, \init, \trans \rangle
}
\newcommand{\sysPair}[2]{
  \langle #1, #2 \rangle
}
\newcommand{\agContract}{C}
\newcommand{\ass}{A}
\newcommand{\gua}{G}
\newcommand{\agTuple}[1]{
  \langle \ass_{#1}[\s{#1}],\allowbreak \gua_{#1}[\s{#1}] \rangle
}
\newcommand{\satisfies}{\models_\mathrm{I}}

\newcommand{\proc}[1]{\ensuremath{\mathsf{#1}}}

\newcommand{\ivcuc}{\textsc{IVC\_UC}\xspace}

\newcommand{\ivcucbf}{\textsc{IVC\_UCBF}\xspace}
\newcommand{\umivc}{\textsc{UMIVC}\xspace}

\newcommand{\mvar}[1]{\mathit{#1}}

\newtheorem{mydef}{Definition}
\newtheorem{prop}{Proposition}

\begin{document}

\title{Merit and Blame Assignment with Kind 2\thanks{
This work was partially funded by DARPA grant \#N66001-18-C-4006 
and by 
GE Global Research.
}}

\author{
  Daniel Larraz\inst{1}
  \and
  Micka{\"e}l Laurent\inst{1}\inst{2}
  \and
  Cesare Tinelli\inst{1}
}
\authorrunning{D. Larraz et al.}

\institute{
Department of Computer Science, The University of Iowa. USA
\and
IRIF, CNRS --- Universit{\'e} de Paris, France
}

\maketitle

\begin{abstract}
We introduce two new major features of the open-source model checker Kind 2
which provide traceability information between specification and design elements
such as assumptions, guarantees, or other behavioral constraints 
in synchronous reactive system models.
This new version of Kind 2 can identify minimal sets of design elements,
known as \emph{Minimal Inductive Validity Cores}, which are sufficient to prove 
a given set of safety properties, 
and also determine the set of \emph{MUST} elements, 
design elements that are necessary to prove the given properties. 
In addition, Kind 2 is able to find minimal sets of design constraints, 
known as \emph{Minimal Cut Sets}, whose violation leads the system 
to an unsafe state. 
The computed information can be used for several purposes, 
including assessing the quality of a system specification,
tracking the safety impact of model changes, and analyzing the tolerance and
resilience of a system against faults or cyber-attacks. 
We describe these new capabilities in some detail and 
report on an initial experimental evaluation of some of them.

\end{abstract}

\section{Introduction}
\label{sec:introduction}

\kind~\cite{Kind2CAV16} is an SMT-based model checker for safety properties of finite- and 
infinite-state synchronous reactive systems. It takes as input models written 
in an extension of the Lustre language~\cite{Lustre92} that allows the specification of 
assume-guarantee-style contracts for system components. 
\kind's contract language~\cite{CoCoSpec16} is expressive enough to allow one 
to represent any (LTL) regular safety property by recasting it 
in terms of invariant properties.
One of \kind's distinguishing features is its support for modular
and compositional analysis of hierarchical and multi-component systems.
\kind traverses the subsystem hierarchy bottom-up, analyzing each system component,
and performing fine-grained abstraction and refinement of the sub-components.
At the component level, \kind runs concurrently several model checking engines
which cooperate to prove or disprove contracts and properties. 
In particular, it combines
two induction-based model checking techniques, $k$-induction~\cite{SheeranSS00} and
IC3~\cite{Bradley11}, with various auxiliary invariant generation methods.

One clear strength of model checkers is their ability to return precise
error traces witnessing the violation of a given safety property.
In addition to being invaluable to help identify and correct
bugs, error traces also represent a checkable unsafety certificate.
Similarly, many model checkers are currently able to return
some form of corroborating evidence when they declare a safety
property to be satisfied by a system under analysis. 
For instance, \kind can produce
an independently checkable proof certificate for the properties that 
it claims to have proven~\cite{MebsoutT16}. However, these certificates, 
in the form of a $k$-inductive invariant, give limited user-level
insight on what elements of the system model contribute to
the satisfaction of the properties.
\medskip

\noindent
\textbf{Contributions}
We describe two new features of \kind 
that provide more insights on verified properties: 
(1) the identification of minimal sets of model elements that are 
sufficient to prove a given set of safety properties, as well as
the subset of design elements that are necessary to prove the
given properties;
(2) the computation of minimal sets of design constraints whose violation leads 
the system to falsify one of more of the given properties.
\medskip

Although these pieces of information are closely related, as we explain later, 
each of them can be naturally mapped to a typical use case in model-based 
software development: respectively,
\emph{merit assignment} and \emph{blame assignment}.
With the former the focus is on assessing the quality of a system specification,
tracking the safety impact of model changes, and assisting in the synthesis of
optimal implementations. With the latter, the goal is to determine
the tolerance and resilience of a system against faults or cyber-attacks.

In general, proof-based traceability information can be used to perform 
a variety of engineering analyses,
including vacuity detection~\cite{VacuityLTL03};
coverage analysis~\cite{ChocklerKP10,ProofBasedMetrics17};
impact analysis~\cite{CompleteTraceability16}, design optimization;
and robustness analysis~\cite{Verdict19,SafetyAnnex20}.
Identifying which model elements are required for a proof,
and assessing the relative importance of different model elements is critical 
to determine the quality of the overall model (including its assume-guarantee
specification), determining when and where to implement changes, 
identifying components that need to be reverified, and measure
the tolerance and resilience of the system against faults and attacks.

\section{Preliminaries}
\label{sec:preliminaries}

Lustre is a synchronous dataflow language that allows one to define system components 
as \define{nodes},
each of which maps a continuous stream of inputs 
(of various basic types, such as Booleans, integers, and reals)
to continuous streams of outputs based on both current input values and 
previous input and output values.
Bigger components can be built by parallel composition of smaller ones, achieved 
syntactically with \define{node applications}. 
Operationally, a node has a cyclic behavior: at each tick $t$ of a global clock
(or a local clock it is explicitly associated with)
it reads the value of each input stream at position or \define{time} $t$, and
instantaneously computes and returns the value of each output stream at time $t$. 

The behavior of a Lustre node is specified declarative by a set of stream constraints
of the form $x = s$, where $x$ is a variable denoting an output or a locally
defined stream and $s$ is a stream algebra over input, output, and local variables.
Most stream operators are point-wise liftings of the usual operators over stream
values.
For example, if $x$ and $y$ are two integer streams, 
the expression $x + y$ is the stream corresponding the function 
$\lambda t. x(t) + y(t)$ over time $t$; 
an integer constant $c$, denotes the constant function $\lambda t. c$. 
Two important additional operators are a unary right-shift operator \code{pre}, used
to specify stateful computations, and a binary initialization operator \code{->},
used to specify initial state values. 
At time $t=0$, the value $(\code{pre}\ x)(t)$ is undefined; 
for each time $t > 0$, it is $x(t-1)$.
In contrast, the value $(x \ \code{->}\ y)(t)$ equals $x(t)$ for $t = 0$
and $y(t)$ for $t > 0$.
Syntactic restrictions guarantee that all streams in a node are inductively well defined. 
In Kind 2's extension of Lustre, nodes can be given assume-guarantee contracts,
enabling the compositional analysis of Lustre models.
Contracts specify assumptions as Boolean terms over current values of input streams and 
previous values of input and output streams, and 
guarantees as Boolean terms over current and previous values of input and output streams.

After various transformations and slicing, \kind 
encodes Lustre nodes internally as state transition systems $\sys=\sysTuple$ 
where $\s{}$ is  a vector of typed state variables, 
$\init$ the initial state predicate, and 
$\trans$ is a two-state transition predicate 
(with $\s{}'$ being a renamed version of $\s{}$).
We will use
$\sysPair{\init}{\trans}$ to refer to transition system $\sys$
when the vector of state variables $\s{}$ is clear from the context or not important. 
A \define{state property} $P$ for a system $\sys=\sysTuple$, 
expressed as a predicate over the variables $\s{}$,
is \define{invariant} if it holds in every reachable state of $S$. 
We will say that $\sys$ \define{satisfies} $P$, written as $\sys \satisfies P$,
if $P$ is invariant for $\sys$.
A \define{contract} for $\sys$ is a pair $\agContract = \agTuple{}$ 
where, informally, the \define{assumption} predicate $\ass$ describes properties
that $\sys$ expects its inputs to have at all times, 
while the \define{guarantee} predicate $\gua$ expresses 
how the component behaves when $\ass$ does hold at all times. 

In \kind, verifying that 
$\sys$ satisfies its contract reduces in essence to verifying 
that $\gua[\s{}]$ is invariant 
for the system $\sys_\ass = \langle\s{}, \init[\s{}] \land \ass[\s{}], \ass[\s{}]
\land \trans[\s{}, \nxtS{}] \land \ass[\nxtS{}]\rangle$.
For hierarchical and multi-component systems, \kind translates the system model
into a hierarchy of transition systems, where a call to a node $N_B$ from a
node $N_A$ is represented by the assertion of the initial (transition) predicate of 
$N_B$ in the initial (transition) predicate of $N_A$ using the correspoding instantiation
of the formal parameters.
Transition systems
provide then a convenient framework not only to check invariant properties
but also to map and refer to different high-level specification and 
design elements in a uniform way. Given a transition system $\sysPair{\init}{\trans}$,
we will assume that $\trans$ has the structure of a top-level conjunction, that is,
$\trans[\s{}, \nxtS{}] = \trans_1[\s{}, \nxtS{}] \land \ldots \land \trans_n[\s{}, \nxtS{}]$
for some $n\geq 1$. Notice that this is the norm in Lustre models where the modeled system
is expressed as the synchronous product of several subcomponents, each of which
is in turn formalized as the conjunction of one or more equational constraints.
Also, Kind 2's assume-guarantee contracts follow naturally this kind of conjunctive
structure since they are specified as conjunction of assumptions and a conjunction
of guarantees. By abuse of notation, we will identify $\trans$
with the set $\{\trans_1, \ldots, \trans_n\}$ of its top-level conjuncts.

\section{Running Example}
\label{sec:example-1}

We will use a simple model to illustrate the concepts and the functionality of \kind
introduced in this report. 
Suppose we want to design a component for an aircraft
that controls the pitch motion of the vehicle, 
and suppose one of the system requirements is that
the aircraft should not ascend beyond a certain altitude. 
The controller must read the current altitude of the aircraft from a sensor, 
and modify the next position of the aircraft's nose accordingly. 
For the sake of simplicity, we will ignore other relevant signals
that should be considered in a real setting to control the elevation of the aircraft. 
Following a model-based design, we model an abstraction of the system's 
environment to which the aircraft's controller will react. 
We also model the fact that the system relies on a possibly imperfect reading 
of the current altitude by an altimeter sensor to decide the next pitch value. 
Finally, we provide a specification for the controller's behavior 
so that it satisfies the system requirement of interest.

\begin{figure}[t]
\begin{lustreTinyMath}[4ex]
node SystemModel (const THRESH, DELTA, S_ERROR: real;
                  sensor_alt: real) returns (actual_alt: real);
(*@contract
  assume "C1: THRESH is positive" THRESH > 0.0; |\label{aC1}|
  assume "C2: DELTA is positive" DELTA > 0.0; |\label{aC2}|
  assume "C3: S_ERROR is non-negative" S_ERROR >= 0.0; |\label{aC3}|
  assume "S: The error in the measured altitude is bounded" |\label{aS}|
    abs(0.0 -> pre actual_alt - sensor_alt) <= S_ERROR;
  guarantee "R1: Altitude is never above THRESH" actual_alt <= THRESH; |\label{gR1}|
*)
  var pitch: real;
let
  pitch = Controller(THRESH, DELTA, S_ERROR, sensor_alt);
  actual_alt = Environment(DELTA, pitch);
tel

node imported Controller (const THRESH, DELTA, S_ERROR: real; 
                          alt: real) returns (pitch: real);
(*@contract
  const LIMIT: real = THRESH - (DELTA + S_ERROR);
  guarantee "L1: Pitch is negative whenever altitude value is over LIMIT"
    alt > LIMIT => pitch < 0.0;
*)
\end{lustreTinyMath}
\caption{System model and subcomponents.
Operators \code{abs} and \code{=>} are respectively the absolute value function and 
Boolean implication.
}
\label{fig:system}
\end{figure}

Our model is described in Figure~\ref{fig:system} in Kind 2's input language.
The main node, \code{SystemModel}, is an \define{observer} node 
that represents the full system consisting in this case of just two components:
one modeling the controller and a node modeling the environment.
The observer has an input \code{sensor\_alt} representing
the altitude value from the altimeter and an output \code{actual\_alt} 
representing the current altitude of the aircraft,
which we are modeling as a product of the environment in response to the pitch value
generated by the controller. 
Of course, in an actual aircraft that value would be communicated to an actuator, 
such as an elevator. 
We are exposing it directly in the observer's interface to simplify the specification 
since we are only interested in the relationship between the actual altitude and 
the pitch value.

\kind allows the user to specify contracts for individual nodes,
either as special Lustre comments added directly inside the node declaration,
or as the instantiation of an external stand-alone contract 
that can be imported in the body of other contracts.
The contract of \code{SystemModel}, included directly in the node, 
specifies assumptions on the altitude value provided by the sensor and 
on a number of symbolic constants 
(\code{THRESH}, \code{DELTA} and \code{S\_ERROR})
which act in effect as model parameters. 
The contract assumes on lines \ref{aC1}--\ref{aC3}
of Figure~\ref{fig:system} 
that those constants are positive---or non-negative for \code{S\_ERROR}. 
The assumption on line~\ref{aS} accounts for fact that, 
while the altitude value produced by the altimeter (\code{sensor\_alt}) 
is not 100\% accurate in actual settings,  
its error is bounded by a constant (\code{S\_ERROR}).
The contract includes a guarantee (on line~\ref{gR1}) that formalizes the requirement 
that aircraft maintain its altitude below a certain threshold \code{TRESH} at all times.
The body of \code{SystemModel} is simply the parallel composition
of the controller component with the environment node.

We do not specify the body of the \code{Controller} and the \code{Environment} nodes
in our model because their details are not important for our purposes.
Instead, we abstract their dynamics with an assume-guarantee contract
that captures the relevant behavior. 
In the \code{Controller}'s case, we model the guarantee that 
the controller will produce a negative pitch value 
whenever the sensor altitude indicates that the aircraft is getting too close 
to the threshold value \code{THRESH} --- with ``too close'' meaning
that the difference between the current altitude and
the threshold is smaller than \code{DELTA + S\_ERROR}
where \code{DELTA} represents an upper bound on the change in altitude
from one execution step to the next (see below).

\begin{figure}[t]
\begin{lustreTinyMath}[4ex]
node imported Environment (const DELTA: real;
                           pitch: real) returns (alt: real);
(*@contract
  guarantee "E1: Altitude is zero initially"
    (alt = 0.0) -> true;
  guarantee "E2: Altitude is always non-negative"
    alt >= 0.0;
  guarantee "E3: Altitude does not increase whenever the controller outputs a negative pitch value"
    true -> (pitch < 0.0 => alt <= pre alt);
  guarantee "E4: Altitude does not decrease more than DELTA units per sampling frame" 
    true -> (pitch < 0.0 => alt >= pre alt - DELTA);
  guarantee "E5: Altitude does not decrease whenever the controller outputs a positive pitch value"
    true -> (pitch > 0.0 => alt >= pre alt);
  guarantee "E6: Altitude does not increase more than DELTA units per sampling frame"
    true -> (pitch > 0.0 => alt <= pre alt + DELTA);
  guarantee "E7: Altitude remains the same whenever the controller outputs a zero pitch value"
    true -> (pitch = 0.0 => alt = pre alt);
*)
\end{lustreTinyMath}
\caption{Contract specification for the \code{Environment} component of \code{SystemModel}.}
\label{fig:environment}
\end{figure}


The declaration of the \code{Environment} component and its contract are shown 
separately in Figure~\ref{fig:environment}.
With \code{alt} representing the actual altitude of the aircraft,
the contract's guarantees capture salient constraints on the physics of our model 
by specifying that a positive pitch value (which has the effect of raising 
the nose of the aircraft and lowering its tail) makes the aircraft ascend,
a negative value makes it descend, and 
a zero value keeps it at the same altitude.\footnote{%
We are ignoring here that, in reality, altitude also depends on aircraft speed.
}
The contract also states that the actual altitude starts at zero, 
is alway non-negative,
and does not change by more than a constant value (\code{DELTA}) in one sampling frame, 
where a sampling frame is identified with one execution step of the synchronous model
for simplicity.\footnote{%
The latter constraint captures physical limitations on the speed of the aircraft.
}

\kind can easily prove that property (guarantee) \code{R1} of \code{SystemModel} is invariant.
However, a few interesting questions arise:

\begin{enumerate}
\item 
Is property \code{R1} satisfied because of the conditions we imposed
on the behavior of \code{Controller}, or does the property trivially hold due 
to the stated assumptions over the environment and the sensor?
\item 
Are all the assumptions over the environment and the sensor in fact necessary 
to prove the satisfaction of property \code{R1}?
\item 
How resilient is the system against the failure of one or more assumptions?
\end{enumerate}

We present the new features of \kind that help us answer these questions next.

\section{The New Features}
\label{sec:functionality}

The first of the two new features offered by \kind consists in identifying 
which parts of the input model were used to construct an inductive proof
of invariance for \code{R1}. 
The new functionality relies on the concept of
inductive validity core introduced by Ghassabani et al.~\cite{IVC16}.
For the rest of the section, let us fix for convenience 
a transition system  $\sys=\sysPair{\init}{\trans}$ and 
an invariant property $P$ for $S$. 

\begin{mydef}
A subset $C \subseteq \trans$ is an \define{inductive validity core} (IVC) for $P$ 
if $\sysPair{\init}{C} \satisfies P$.
\end{mydef}

Note that, as $P$ is invariant for $S$ (i.e., $\sysPair{\init}{\trans} \satisfies P$),
$\trans$ is already an IVC, although not a very interesting one. 
In practice, it is often possible to compute efficiently 
a smaller IVC that contains fewer or no irrelevant elements
(see Section~\ref{sec:implementation} for more details).
We can ensure that the elements of an IVC for a property $P$ are necessary 
by requiring it to be \define{minimal}, that is, 
to have no proper subset that is also IVC for $P$.

One can see every IVC as an \define{approximate} minimal IVC (MIVC).
Computing approximate MIVCs is the recommended option in \kind
for quickly detecting modeling issues such as various forms of vacuity 
due to inconsistent assumptions or transition relation.
\kind can compute efficiently a single MIVC for a property $P$
(see Section~\ref{sec:implementation} for a brief description 
of the method, based on one by Ghassabani et al.~\cite{IVC16})
and offers the option to compute all MIVCs for $P$.

\subsection{IVCs for coverage and change impact analysis}

If a property $P$ of a system $S$ has multiple MIVCs, inspecting all of them 
provides insights on the different ways $S$ satisfies $P$. 
Moreover, given all the MIVCs for $P$,
it is possible to partition all the model elements into three sets: 
a \emph{MUST} set of elements which are required for proving $P$ in every case, 
a \emph{MAY} set of elements which are optional, and 
a set of elements that are irrelevant.
More formally, let $\mvar{MIVCs}(S,P)$ denote the set of all MIVCs
for $S$ and $P$. 
Then, we have the following categorization~\cite{CompleteTraceability16}:

\begin{itemize}
\item $\mvar{MUST}(S,P) = \bigcap \mvar{MIVCs}(S,P) $ 
\item $\mvar{MAY}(S,P) = (\bigcup \mvar{MIVCs}(S,P)) \setminus \mvar{MUST}(S,P)$
\item $\mvar{IRR}(S,P) = \trans \setminus (\bigcup \mvar{MIVCs}(S,P))$
\end{itemize}

This categorization provides complete traceability between specification and
design elements, and can be used for coverage analysis~\cite{ProofBasedMetrics17} and 
tracking the safety impact of model changes.
For instance, a change to one of the elements in $\mvar{MAY}(S,P)$ will not affect the 
satisfaction of $P$ but will definitely impact some other property $Q$ 
if it occurs in $\mvar{MUST}(S,Q)$.
Furthermore, we can use the set $\mvar{IRR}(S,P_1\land\ldots\land P_m)$ 
of irrelevant elements for the conjunction of all the invariant properties 
$P1,\ldots,P_m$ of $S$ to determine their completeness: 
if the set is non-empty that indicates that there may be missing requirements.

\begin{example}
If we ask \kind to generate an approximate MIVC for the invariant \code{R1}
of the system presented in Section~\ref{sec:example-1},
\kind generates a IVC with 7 elements:
assumptions \code{S} and \code{C1} from \code{SystemModel}'s contract,
the (only) guarantee \code{L1} in \code{Controller}'s contract, and 
all guarantees in \code{Environment}'s contract except 
for \code{E2}, \code{E4}, and \code{E5}.

This tells us already that \code{E2}, \code{E4}, and \code{E5} are not necessary
to satisfy property \code{R1} and is enough to answer the second of the questions 
listed at the end of Section~\ref{sec:example-1}. 
Moreover, since the guarantee \code{L1} of \code{Controller} is part of the IVC, 
it is likely that the controller's behavior is relevant for the satisfaction 
of \code{R1}. 
However, we can not be sure because the generated IVC is an \emph{approximate} MIVC
and so is not necessarily minimal. 

To confirm that \code{L1} is indeed necessary we can ask \kind to identify a true MIVC, 
a more expensive task computationally.
When we do that, \kind returns the same set.
This confirms the necessity of the guarantee \code{L1} 
but only for the specific proof of \code{R1}'s invariance found by Kind 2.
It might still be the case that the guarantee is not required in general,
that is, there may be \emph{other} proofs that do not use \code{L1}, 
which would be confirmed by the discovery of a different MIVC 
that does not contain it.
In other words, at this point we do not know 
whether \code{L1} is a \emph{must} element for \code{R1}.
To determine that, and also to know if the assumptions not included 
in the computed MIVC are always irrelevant for the satisfaction of \code{R1},
we can ask \kind to compute \emph{all} MIVCs.
In that case, \kind will return only one MIVC, the set found in the first analysis,
which confirms that all the included elements are required and 
the excluded ones are irrelevant.
In alternative, we could have asked \kind to compute a single MIVC \emph{and}
the MUST set for property \code{R1}. As we show later in this report,
\kind uses a method for generating the MUST set that
does not require the computation of all MIVCs.
\qed
\end{example}

\subsection{IVCs for fault-tolerance or cyber-resiliency analysis}

Another use of IVCs, and  MUST sets in particular, is in the analysis 
of a system's tolerance to faults~\cite{SafetyAnnex20} or 
resiliency to cyberattacks~\cite{Verdict19}.
For instance, an empty MUST set for a system $S$ and its invariant $P$ indicates 
that the property is satisfied by $S$ in various ways, 
making the system fault tolerant or resilient against cyberattacks 
as far as property $P$ is concerned.
In contrast, a large MUST set, as in our running example, suggest a more brittle system, 
with multiple points of failure or a big attack surface.

\begin{figure}[t]
\begin{lustreTinyMath}[4ex]
node SystemModel (const THRESH, DELTA, S_ERROR: real;
                  alt1, alt2, alt3: real) returns (actual_alt: real);
(*@contract
  assume "C1: THRESH is positive" THRESH > 0.0;
  assume "C2: DELTA is positive" DELTA > 0.0;
  assume "C3: S_ERROR is non-negative" S_ERROR >= 0.0;
  assume "S1: Error in altitude from sensor 1 is bounded by S_ERROR"
    abs(0.0 -> pre actual_alt - alt1) <= S_ERROR;
  assume "S2: Error in altitude from sensor 2 is bounded by S_ERROR"
    abs(0.0 -> pre actual_alt - alt2) <= S_ERROR;
  assume "S3: Error in altitude from sensor 3 is bounded by S_ERROR"
    abs(0.0 -> pre actual_alt - alt3) <= S_ERROR;
  guarantee "R1: Altitude never above THRESH" actual_alt <= THRESH;
*)
  var pitch, alt: real;
let
  alt = TriplexVoter(alt1, alt2, alt3);
  pitch = Controller(THRESH, DELTA, S_ERROR, alt);
  actual_alt = Environment(DELTA, pitch);
tel
\end{lustreTinyMath}
\caption{Enhanced system model.}
\label{fig:system-enhanced}
\end{figure}

\begin{figure}[t]
\begin{lustreTinyMath}[4ex]
node TriplexVoter (alt1,alt2,alt3: real) returns (r: real);
  var ad12,ad13,ad23,m: real;
let
  ad12 = abs(alt1 - alt2);  ad13 = abs(alt1 - alt3);
  ad23 = abs(alt2 - alt3);

  m = min(ad12, min(ad13, ad23));

  r = if m = ad12 then (alt1 + alt2) / 2.0
      else if m = ad13 then (alt1 + alt3) / 2.0
      else (alt2 + alt3) / 2.0;
tel
\end{lustreTinyMath}
\caption{Low-level specification of the Triplex voter.
Operator \code{min} computes the minimum of its two inputs.}
\label{fig:triplex-voter}
\end{figure}

\begin{example}
Our previous analysis on the system model of our running example 
confirms that the correctness of the altitude sensor is crucial 
for the system not to exceed the prescribed altitude limit. 
One way to improve the system fault-tolerance then is 
to introduce some redundancy.
In particular, we could equip the system with three different altimeters and
so send to the controller three independent altitude values.
The controller, or a dedicated new subcomponents could then take the average 
of the two altitude values that are closest to each other 
(as they more likely to be close to the actual altitude).
This way, if one of the altimeter fails, in the sense that it produces 
an altitude reading with an error greater than the maximum expected error, 
the other two values allows the system to compensate for that error. 
We can easily change the model to implement this redundancy mechanism. 
First, we extend the interface of \code{SystemModel} to take three altitude values 
(\code{alt1}, \code{alt2}, and \code{alt3}) instead of one, and then 
we introduce a new component, a triplex voter 
that takes those sensor values and computes an estimated altitude for the controller
as explained above.
The new specification for \code{SystemModel} is provided 
in Figure~\ref{fig:system-enhanced} with an updated contract for \code{SystemModel} 
that independently assumes the same error bounds on each individual altitude value.
A full specification for the triplex voter is given 
in Figure~\ref{fig:triplex-voter}.
\medskip

We can use \kind to confirm that property \code{R1} still holds
after the introduction of the two new sensors and the triplex voter.
However, this result is not enough to determine whether the introduced redundancy 
mechanism makes the system more fault tolerant. 
To confirms that we must verify the existence of multiple MIVCs. 
Perhaps surprisingly though, when we ask \kind to compute all the MIVCs, 
it still reports a single solution---which includes all the sensor assumptions.
Put differently, the MIVC analysis, shows that 
the satisfaction of property \code{R1} requires \emph{all three} sensors 
to behave accordingly to their specification. 
After reviewing the model, however, one can conclude 
that to benefit from the triplex voter it is necessary to decrease 
the safety limit value \code{LIMIT} in the controller's contract. 
In particular, it is enough to decrease it as follows, 
doubling the error bound value:
\smallskip

\begin{lustreTinyMath}[4ex]
  const LIMIT: real = THRESH - (DELTA + 2.0*S_ERROR);
\end{lustreTinyMath}
\smallskip

After this change, the number of MIVCs reported by \kind increases 
from one to three. 
Each MIVC contains two of the assumptions \code{S1}, \code{S2}, \code{S3} 
on the three sensors, and the rest of the assumptions and guarantees included 
in the MIVC computed for the previous version of the model.
This illustrates how the new traceability feature in \kind could be used
to detect a subtle flaw in the enhanced model that prevents it 
from making the system fault-tolerant despite the triplication of the altitude sensors.
We stress how a simple safety analysis, verifying the invariance of \code{R1} would not
help detect such flaw.
\qed
\end{example}

\subsection{Quantifying a system's resilience}

To help quantify the resilience of a system, \kind also supports 
the computation of minimal cut sets (aka, \emph{minimal correction sets})
for an invariance property.

\begin{mydef} 
A subset $C \subseteq \trans$ is a \define{cut set} for $P$ if
$\sysPair{\init}{\trans \setminus C} \not\satisfies P$. 
A \define{minimal cut set} (MCS) for $P$ is a cut set 
none of whose proper subsets is a cut set for $P$. 
A \define{smallest cut set} is an MCS of minimum cardinality. 
We will use $\mvar{MCSs}(S,P)$ to denote the set of all MCS for $S$ and $P$.
\end{mydef}

\kind provides options to compute a (single) smallest cut set, all the MCSs, and
all the MCSs up to a given cardinality bound. 
In the context of fault or security analyses, 
the cardinality of an MCS for $P$ represents 
the number of design elements that must fail or be compromised 
for the property to be violated.
The smaller the MCS, or the higher the number of MCSs of small cardinality,
the greater the probability that the property can be violated.

It is worth mentioning that there exists a hitting set duality between MCSs
and MIVCs, reflected by the following proposition, which provides a method to compute 
all MCSs from the set of all MIVCs:

\begin{prop}[Theorem 1 in~\cite{ChasingIVC19}]
A subset $C \subseteq \trans$ is an MIVC for $P$
iff $C$ is a minimal hitting set of $\mvar{MCSs}(S,P)$. 
\end{prop}

Notwithstanding the result above, there is a more direct approach,
implemented in \kind and described in Section~\ref{sec:implementation},
for computing all the MCSs which has the benefit of generating 
MCSs incrementally, without having to wait for the computation of all MIVCs.
Similarly, 
there is a method for generating $\mvar{MUST}(S,P)$ that does not require 
the computation of all MIVCs.
The method is based on the observation that 
$\mvar{MUST}(S,P)$ consists of the union of all MCSs of cardinality 1 
for $S$ and $P$.

\begin{prop}
\label{prop:MUST-from-MCSs1}
$\mvar{MUST}(S,P)=\{e \mid \{e\}\in \mvar{MCSs}(S,P)\}$.
\end{prop}

\kind offers the option to compute the MUST set together with a single MIVC or all of them. 
As the experiments of Section~\ref{sec:evaluation} show, 
the overhead of computing the MUST set in addition to an MIVC is negligible. 
Furthermore, \kind uses the computation of the MUST set to check whether there exists
only one MIVC, and terminate the search for more MIVCs early. 
The check is based on the 
result.

\begin{prop}
\label{prop:MUST-is-IVC}
If $\mvar{MUST}(S,P)$ is an IVC then 
it is minimal and unique, i.e., $\mvar{MIVCs}(S,P)=\{\mvar{MUST}(S,P)\}$.
\end{prop}

When looking for all MIVCs, 
\kind always computes the set of MUST elements first and
then checks whether $\mvar{MUST}(S,P)$ is enough to prove $P$. 
If that is the case, it stops looking for more MIVCs. 
The experiments in Section~\ref{sec:evaluation} indicate
that this is an effective strategy.

\begin{example}
If we ask \kind to compute all the MCSs for the version of the running
example that includes the redundancy mechanism, \kind finds the following
MCSs: $\{\code{E1}\}$, $\{\code{E3}\}$, $\{\code{E6}\}$, $\{\code{E7}\}$,
$\{\code{C1}\}$, $\{\code{L1}\}$, $\{\code{S1},\code{S2}\}$,
$\{\code{S1},\code{S3}\}$, $\{\code{S2},\code{S3}\}$.

\end{example}

\section{Implementation details}
\label{sec:implementation}

We now give a high-level description of the algorithms implemented in \kind
to provide the functionality described in Section~\ref{sec:functionality}.
Again, we fix a transition system $\sys = \sysPair{\init}{\trans}$ 
with $\trans = \{\trans_1, \ldots, \trans_n\}$ and 
an invariant property $P$ for $\sys$.

The main algorithms use a number of auxiliary procedures,
namely, \textsf{Verify}, \textsf{MinimizeIVC},
\textsf{GetMCS}, and \textsf{GetApproximateMIVC},
which are non-terminating in general for being based on checking 
the invariance of state properties of transition systems,
an undecidable problem in the infinite-state case.
In our concrete implementation, we make the main algorithms terminating 
by imposing a time limit on these procedures and handling timeout exceptions 
as follows.
For \textsf{Verify}, we extend the type of the returned result
with an additional value, \texttt{unknown}, to account for the exception.
For \textsf{MinimizeIVC}, \textsf{GetMCS}, and the main algorithms, 
we have them return also a Boolean value indicating 
whether the result is precise or approximate due to a timeout.
For \textsf{GetApproximateMIVC}, which calls \textsf{Verify} and
then uses algorithm \ivcuc from~\cite{IVC16},
we have the procedure simply return $T$ if \textsf{Verify}
returns \texttt{unknown}, or the result of \ivcuc otherwise.


\subsection{Computing approximate MIVC and single MIVC}

The computation of an approximate MIVC is based on algorithm \ivcuc 
by Ghassabani et al.~\cite{IVC16}.
It consists of three main steps:
$(i)$ 
reducing the value of $k$ for the $k$-inductive proof of property $P$
(obtained by finding a k-inductive strengthening 
$Q = Q_1 \land \ldots \land Q_n$ of $P$); 
$(ii)$ 
reducing the number of conjuncts in invariant $Q$ 
by removing those not needed in the proof; 
$(iii)$
computing an UNSAT core over the model constraints in the same query
to the backend SMT solver 
that checks that $Q$ is a k-inductive strengthening of $P$.

The computation of a single MIVC is based on algorithm \ivcucbf, also
by Ghassabani et al.~\cite{IVC16}.
The main idea is to generate an \define{approximate} MIVC first, 
and then minimize it using a brute-force approach
that removes one model element at a time and (model) checks 
that the property $P$ still holds.

\subsection{Computing all MIVCs}

\begin{algorithm}[t]
\begin{algorithmic}[1]
  \State $\mvar{MIVCs}$ := $\emptyset$; $\mvar{approx}$ := $\texttt{false}$
  \State $\mvar{MUST}, \mvar{ap}$ := \Call{MUST-Set}{$\s{}$, $S$, $P$} \label{ivc:line:MUST-set}
  \State $A$ := $\{a_i\mid 1\leq i\leq|T|\}$ \Comment{Fresh bool variables}
  \State $\mvar{map}$ := $\bigwedge_{T_i\in \mvar{MUST}} a_i$ \label{ivc:line:map-init}
  \State $\mvar{res}, \theta$ := \proc{Verify}($I$, $\mvar{MUST}$, $P$)
  \If{$\mvar{res}$=$\texttt{safe}$} \label{ivc:line:MUST-is-IVC}
    \State $\mvar{MIVCs}$ := $\{\mvar{MUST}\}$
  \Else
    \While{\textsf{IsSAT}($\mvar{map}$)} \label{ivc:line:while-head}
      \State $\mvar{seed}$ := \proc{GetUnexploredMax}($\mvar{map}$, $T$) \label{ivc:line:extract-seed}
      \State $\mvar{res}, \theta$ := \proc{Verify}($I$, $\mvar{seed}$, $P$)
      \If{$\mvar{res}$=$\texttt{safe}$} \Comment{$\mvar{seed}$ is an IVC}
        \State $\mvar{mivc}, \mvar{ap}$ := \proc{MinimizeIVC}($\mvar{seed}$, $\mvar{MUST}$, $S$, $P$) \label{ivc:line:minIVC}
        \State $\mvar{MIVCs}$ := $\mvar{MIVCs} \cup \{(\mvar{mivc}, \mvar{ap})\}$ \label{ivc:line:solution}
        \State $\mvar{map}$ := $\mvar{map} \land \bigvee_{T_i\in \mvar{mivc}} \neg a_i$ \label{ivc:line:blockMIVC}
      \Else
        \State $\mvar{mcs}$, $\mvar{ap}$ := $\proc{GetMCS}(S, T \setminus \mvar{seed}, P)$ \label{ivc:line:getMCS}
        \State $\mvar{map}$ := $\mvar{map} \land \bigvee_{T_i\in mcs} a_i$ \label{ivc:line:blockMCS}
        \State $\mvar{approx}$ := $\mvar{approx} \lor \mvar{res}=\texttt{unknown}$ \label{ivc:line:unknown}
      \EndIf
    \EndWhile \label{ivc:line:while-end}
  \EndIf
  \State \Return $(\mvar{MIVCs}, \mvar{approx})$
\end{algorithmic}
\caption{AllMIVCs($S=\sysTupleAbr$, $P$)}
\label{alg:AllMIVCs}
\end{algorithm}

To compute all MIVCs for $\sys$ and $P$
we adapted algorithm \umivc by Berryhill and Veneris~\cite{ChasingIVC19}
which in turn is a generalization of previous work~\cite{All-IVC-Online18,All-IVC-17}.
Our version, described in Algorithm~\ref{alg:AllMIVCs},
starts by generating $\mvar{MUST}(S,P)$ (line~\ref{ivc:line:MUST-set}) 
which is then used (line~\ref{ivc:line:map-init}) to provide an initial value
for variable $\mvar{map}$.
That variable stores a CNF formula that tracks the portions of 
the power set of $T$ explored so far. 
Each satisfying assignment for $\mvar{map}$ corresponds to
an abstraction of $T$ that includes an element $T_i$ 
exactly if variable $a_i$ is true.
The value of $\mvar{ap}$ in line~\ref{ivc:line:MUST-set},
which indicates $\mvar{MUST}$ is an underapproximation due to a timeout,
is ignored because it does not affect soundness or completeness.
Line~\ref{ivc:line:MUST-is-IVC} checks whether $\mvar{MUST}$ is an IVC. If it is,
there exists only one MIVC and it corresponds to $\mvar{MUST}(S,P)$
(see Proposition~\ref{prop:MUST-is-IVC}). Otherwise, lines 
\ref{ivc:line:while-head}-\ref{ivc:line:while-end} compute all MIVCs.
Line~\ref{ivc:line:extract-seed} extracts an abstraction of $T$ from the map
called $\mvar{seed}$ of maximum cardinality using a MaxSAT solver. 
If the $\mvar{seed}$ is not an IVC or the result is unknown,
line~\ref{ivc:line:getMCS} computes an MCS over $T \setminus \mvar{seed}$
that is used to prune the search space of candidates in line~\ref{ivc:line:blockMCS}.
The new clause forces any candidate
solution to include at least one element of the MCS.
Treating an unknown result as unsafe is sound, but it may affect completeness.
In that case, $\mvar{approx}$ is set to true in line~\ref{ivc:line:unknown}.
If the $\mvar{seed}$ is an IVC, line~\ref{ivc:line:minIVC} reduces the $\mvar{seed}$ 
to an MIVC using an optimized version of algorithm \ivcucbf that takes into account 
the fact that none of the elements in $\mvar{MUST}$ can be removed.
If the minimization fails, the solution is tagged as approximate
in line~\ref{ivc:line:solution}.
Then, line~\ref{ivc:line:blockMIVC} adds
a clause to $map$ that blocks the MIVC and all its supersets.
The algorithm returns the computed set and a boolean value that is true
if the solution is approximate, that is, there is no guarantee that
every MIVC is contained in some element of the set.

Algorithm~\ref{alg:AllMIVCs} can be seen as an instantiation of \umivc
where all MCSs of cardinality $1$ are precomputed. 
The major difference with \umivc is that our algorithm is able to identify 
the MUST set from the generated set of MCSs
(c.f. Proposition~\ref{prop:MUST-from-MCSs1}), 
and can use it to check for early termination
and to boost the minimization of the IVC in line~\ref{ivc:line:minIVC}.
Our current implementation in \kind is in fact a generalization 
of Algorithm~\ref{alg:AllMIVCs} 
that allows the user to choose the set of model elements 
(assumptions and guarantees, node calls, equations in node bodies, \ldots) 
over which the MIVCs must be computed.
It assumes that the rest of model elements are always present in the system.

\begin{algorithm}[t]
\begin{algorithmic}[1]
  \State $\mvar{MCSs}$ := $\emptyset$; $\mvar{approx}$ := $\texttt{false}$
  \State $m$ := $min(ub,|E|)$
  \State $\y{}=\langle y_1,\ldots,y_{|E|} \rangle$ \Comment{Fresh bool variables}
  \State $T^{\star}$ := $\bigwedge_{T_i\in E} (\neg y_i \Rightarrow T_i \land y_i'=y_i)
          \land \bigwedge_{T_j\in (T \setminus E)} T_j$
  \State $k$ := $m$; $\mvar{res}$ := $\texttt{unknown}$; $\theta$ := $\emptyset$
  \Do \label{mcs:line:do-head}
    \State $I^{\star}$ := $I \land \proc{AtMostK}(\y, k)$
    \State $\mvar{res}, \theta'$ := \textsf{Verify}($I^{\star}$, $T^{\star}$, $P$)
    \If {$\mvar{res}=\texttt{unsafe}$}
      \State $\theta$ := $\theta'$; 
      $k$ := $k - 1$ \Comment{Store last counterexample, and decrement $k$}
    \EndIf
  \doWhile{$k\geq 0 \land \mvar{res}=\texttt{unsafe}$} \label{mcs:line:do-while}
  \State $\mvar{approx}$ := $\mvar{approx} \lor \mvar{res}=\texttt{unknown}$
  \If {$k < m$} \Comment{Unsafe for $k+1$} \label{mcs:line:exists-sol}
    \If {$k < 0$} $\mvar{MCSs}$ := $\{\emptyset\}$ \Comment{No Safe for any $k$} \label{mcs:line:unsafe}
    \Else
       \State $k$ := $k + 1$ \Comment{Cardinality of a smallest MCS}
       \State $C$ := \proc{ExtractCutSet}($\theta$, $\y$, $E$) \label{mcs:line:first-mcs}
       \State $\mvar{MCSs}$ := $\mvar{MCSs} \cup \{(C, \mvar{approx})\}$
       \State $\phi$ := $\bigvee_{T_i\in C} \neg y_i$ \Comment{Block supersets of $C$} \label{mcs:line:block1}
       \While {$k \leq m$} \label{mcs:line:while-head-ext}
         \State $I^{\star}$ := $I \land \proc{AtMostK}(\y, k)$
         \State $\mvar{res}, \theta'$ := \proc{Verify}($I^{\star}\land\phi$, $T^{\star}$, $P$)
         \While {$\mvar{res}=\texttt{unsafe}$} \label{mcs:line:while-head-inn}
           \State $C$ := \proc{ExtractCutSet}($\theta'$, $\y$, $E$)
           \State $\mvar{MCSs}$ := $\mvar{MCSs} \cup \{(C, \mvar{approx})\}$
           \State $\phi$ := $\phi \land \bigvee_{T_i\in C} \neg y_i$
           \State $\mvar{res}, \theta'$ := \proc{Verify}($I^{\star}\land\phi$, $T^{\star}$, $P$)
         \EndWhile \label{mcs:line:while-end-inn}
         \State $\mvar{approx}$ := $\mvar{approx} \lor \mvar{res}=\texttt{unknown}$
         \State $k$ := $k + 1$ \label{mcs:line:increment}
       \EndWhile \label{mcs:line:while-end-ext}
    \EndIf
  \EndIf
  \State \Return $(\mvar{MCSs}, \mvar{approx})$
\end{algorithmic}
\caption{AllMCSs\_UpToUB($S=\sysTupleAbr$, $E$, $P$, $\mvar{ub}$)}
\label{alg:AllMCSs}
\end{algorithm}

\subsection{Computing all MCSs}

Given a subset $E$ of $\trans = \{\trans_1, \ldots, \trans_n\}$,
\kind uses an implementation of Algorithm~\ref{alg:AllMCSs}
to find all the MCSs over $E$ for $\sys$ and $P$  
with cardinality not greater than a given upper bound $ub$.
The algorithm can be used to compute all MCSs 
for $S$ and $P$ by choosing $E$ to be $\trans$ and $ub$ to be $|\trans|$.

The algorithm is based on techniques previously applied to
automated debugging~\cite{FaultDiagnosis19,BerryhillV18}.
To understand how it works it is helpful to observe that
we can reduce the problem of finding one cut set 
for $\sys$ and $P$ to a model checking problem.
Specifically, we build a modified version of $\sys$,
$\sys^{\star}=\langle \z{}, \init[\z{}], \trans^{\star}[\z{}, \z{}'] \rangle$,
where the vector of typed variables $\z{}$ includes $\s{}$ and fresh Boolean
variables $\y{}=\langle y_1,\ldots,y_{|T|} \rangle$, and the transition
predicate is defined by $\trans^{\star}[\z{}, \z{}'] = \bigwedge_{T_i\in T} 
(\neg y_i \Rightarrow T_i \land y_i'=y_i)$.
Then, we check whether $S^{\star}$ satisfies $P$ or not. 

If we can disprove $P$, 
the initial values assigned to $\y{}$ in any trace that leads
$S^{\star}$ to the violation of $P$ determines a cut set:
the set of all $T_i$'s such as the initial value of $y_i$ is true. 
Moreover,
if we add a cardinality constraint over $\y{}$ in $I$ specifying that at most
$k$ of the variables are true, any solution $C$ will be such that $|C|\leq k$.
Lines~\ref{mcs:line:do-head}-\ref{mcs:line:do-while} use this reduction to
find an MCS of minimal cardinality.
If none exists, or the first call to \textsf{Verify} returned unknown,
the condition in 
line~\ref{mcs:line:exists-sol} is false, and the empty set is returned.
If the original system $S$ does not satisfy $P$, condition in
line~\ref{mcs:line:unsafe} is true, and the only MCS is the empty set.
Otherwise, lines~\ref{mcs:line:first-mcs}-\ref{mcs:line:while-end-ext} compute
all (non-empty) MCSs. The first MCS is extracted in
line~\ref{mcs:line:first-mcs} from $\theta$, which is the last error trace found
in the do-while loop. To block the found MCS and all its supersets,
line~\ref{mcs:line:block1} initializes $\phi$, a CNF formula which tracks 
which portion of the power set of $E$ has been explored. 
After that, the rest of MCSs
of cardinality $k$~\footnote{The minimal cardinality 
if $approx$ has not been set to true.}
are extracted in the internal loop
(lines~\ref{mcs:line:while-head-inn}-\ref{mcs:line:while-end-inn}).
When no more MCSs of cardinality $k$ exists, $k$ is incremented by one in
line~\ref{mcs:line:increment}. 
If the maximum cardinality has not been reached yet,
checked in line~\ref{mcs:line:while-head-ext}, the internal loop continues
to extract all the MCSs of cardinality $k$. 
Since $\phi$ contains clauses that block any superset of an MCS of cardinality 
less than $k$, the new MCSs generated are guaranteed to be minimal
if no call to \textsf{Verify} has returned unknown. Otherwise, minimality
is not guaranteed, and the solution is tagged as approximate. Moreover,
the algorithm returns true together the generated set to indicate that
some MCS may not be contained in any element of the set.

\subsection{Computing the MUST set}

\begin{algorithm}[t]
\begin{algorithmic}[1]
  \State $\mvar{ivc}$ := \proc{GetApproximateMIVC}($S$, $P$) \label{must:line:ivc}
  \State $\mvar{MCSs}, \mvar{approx}$ := \Call{AllMCSs\_UpToUB}{$S$, $\mvar{ivc}$, $P$, $1$} \label{must:line:mcs}
  \State \Return $(\{e \mid (\{e\},\texttt{false})\in \mvar{MCSs}\}, \mvar{approx})$ \label{must:line:must}
\end{algorithmic}
\caption{MUST-Set($S=\sysTupleAbr$, $P$)}
\label{alg:MUST-Set}
\end{algorithm}

We use Algorithm~\ref{alg:MUST-Set} to compute the MUST set for a transition system $S$
and an invariant property $P$. 
It first finds an \emph{approximate} MIVC, $\mvar{ivc}$,
using algorithm \ivcuc from~\cite{IVC16}. 
Then, it uses the $\mvar{ivc}$ to narrow the search of MCSs of cardinality $1$.
When the result returned in line~\ref{must:line:mcs} is precise,
it builds the MUST set from the generated MCSs
relying on Proposition~\ref{prop:MUST-from-MCSs1}, and returns the set
together the boolean value false.
Otherwise, the algorithm returns an underapproximation~\footnote{Notice that
every returned cut set (of cardinality 1) is minimal if $P$ is invariant.}
of the MUST set together the boolean value true.
This algorithm shares some similarities with one 
by Chockler et al.~\cite{ChocklerKP10}
for determining non-deterministic mutation coverage (\textsf{\small Nondet-Cov}). 
This is expected since \textsf{\small Nondet-Cov} is equivalent to coverage based 
on MUST elements, as proved by Ghassabani et al.~\cite{ProofBasedMetrics17}.

\section{Experimental Evaluation}
\label{sec:evaluation}

Using the set of benchmarks from~\cite{All-IVC-17}, 
we ran two experiments\footnote{%
All materials and pointers to the new version of \kind are available 
at \url{https://github.com/kind2-mc/mivc-must-experiments}.
}
to address the following questions:
$(i)$ how the computation of both a single MIVC (using \ivcucbf) and the MUST set (using Algorithm~\ref{alg:MUST-Set}) 
compares performance-wise
with the computation of just an MIVC, and
$(ii)$ 
what is the actual advantage of computing the MUST set to prune the search space and 
checking the existence of a single MIVC in Algorithm~\ref{alg:AllMIVCs}.
 
\begin{figure}[t]%
  \centering
\begin{subfigure}[b]{.49\textwidth}%
  \includegraphics[width=\textwidth]{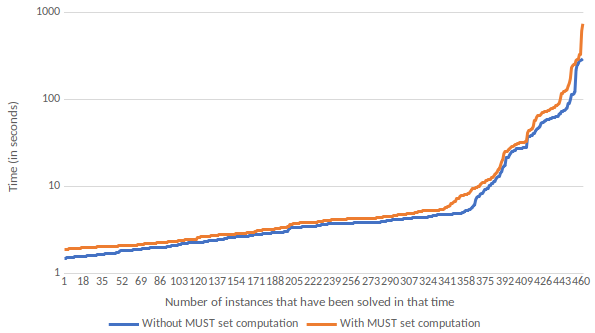}
  \caption{Computation of a single MIVC}
  \label{fig:Q1}
\end{subfigure}
\begin{subfigure}[b]{.49\textwidth}%
  \includegraphics[width=\textwidth]{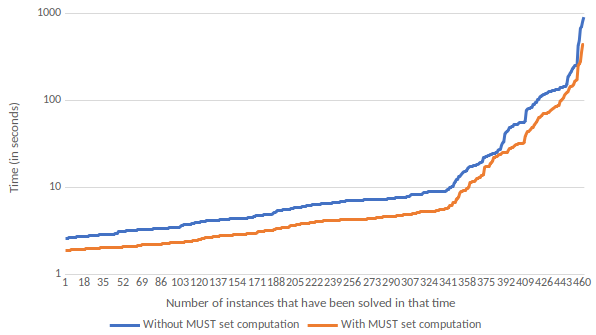}
  \caption{Computation of all MIVCs}
  \label{fig:Q2}
\end{subfigure}
  \caption{Experimental Results}
    \label{fig:results}
\end{figure}

For question $(i)$, our results, illustrated in Figure~\ref{fig:Q1}, 
reveal that the overhead of generating the MUST set in addition to a single MIVC
is negligible. 
However, the feedback provided to the users is highly informative 
as it allows them to quickly check if there is only one MIVC and, 
if not, determine how many elements of the MIVC are required.
For question $(ii)$,
our results confirm that the optimization is beneficial as
Algorithm~\ref{alg:AllMIVCs} exhibits on average a 1.5x speedup
over the version without the computation,
as illustrated by the cactus plot in Figure~\ref{fig:Q2}.

\section{Related Work}

The computation of an approximate MIVC was first available in
the open-source model checker \jkind~\cite{JKind} around
the same time the technique was introduced in~\cite{IVC16}.
More recently, \jkind started to offer support for
the computation of all MIVCs based on
the \emph{offline} algorithm decribed in~\cite{All-IVC-17}.
The algorithm is considered \emph{offline} because it is not 
until all IVCs have been computed that one knows whether the
solutions computed are, in fact, minimal. For
models contain many IVCs, this approach can be impractically expensive
or simply not terminate. However, for applications where only a full
enumeration of the MIVCs is 
this technique may offer better overall performance. The main idea is
to use algorithm \ivcuc for the minimization of the IVC in
line~\ref{ivc:line:minIVC} of Algorithm~\ref{alg:AllMIVCs},
as opposed to the more expensive algorithm \ivcucbf that
ensures minimality, and to not minimize the cut set in
line~\ref{ivc:line:getMCS} before adding it to the map.

Although not part of the official distribution of \jkind,
the \emph{online} algorithm for computing all MIVCs
presented in~\cite{All-IVC-Online18} has also been implemented
in the tool. Similarly to Algorithm~\ref{alg:AllMIVCs},
it incorporates the idea of reducing the cardinality of
the cut sets generated when calls to \textsf{Verify} returns
unsafe. In contrast, the method only tries to reduce
the cardinality when \textsf{Verify} returns unsafe within
algorithm \ivcucbf, not in the main loop of
Algorithm~\ref{alg:AllMIVCs}.
Moreover, the reduction is based on retrieving a maximal
set of \emph{map} that contains the seed,
and checking whether the subset is a IVC or not.
If it is not a IVC, the complement of the subset is
an approximation of a MCS.
Otherwise, approximate MIVCs are computed and used
to reduce the elements of the seed until the subset
is not a IVC anymore.

Unlike \kind, \jkind does not have native support for assume-guarantee
contracts in its input language. Thus, \jkind only considers equations
for the generation of IVCs. In contrast, \kind allows the user to
select 
not only design elements
such as node calls and equations but also specification elements 
such as assumptions and guarantees. 


An algorithm for computing all MCSs is described by Bozzano et al.~\cite{BozzanoCGM15}.
Like our Algorithm~\ref{alg:AllMCSs}, their technique computes the cuts sets 
of increasing cardinality to prevent the generation of non-miminal solutions.
However, their method relies on a IC3-based routine for parameter 
synthesis to compute all the solutions in each layer. Therefore,
instead of relying on a black-box \textsf{Verify} procedure to solve
multiple ordinary model checking queries, they use a specialized algorithm.
The main advantage in that case is that the information learnt to block a particular 
counterexample can be reused when considering new ones.

%

\bibliographystyle{splncs04}
\bibliography{main.bib}

\end{document}